\documentclass[aps,amsmath,amssymb,preprintnumbers,groupedaddress,twocolumn,floatfix]{revtex4}
\usepackage{amsmath,amssymb,epsfig,amsfonts}
\usepackage{graphicx,floatflt,subfigure}
\def\unit{{1\kern-.65ex {\rm l}}}
\def\1{{1\kern-.65ex {\rm l}}}

\newcount\hour \newcount\minute
\hour=\time \divide \hour by 60
\minute=\time
\def\now{%
\ifnum \hour<13
  \ifnum \hour=0 \advance \hour by 12 \number\hour:\else \number\hour:\fi%
     \ifnum \minute<10 0\fi%
     \number\minute%
\ A.M.%
\else \advance \hour by -12 \number\hour:%
  \ifnum \minute<10 0\fi%
  \number\minute%
  \ P.M.%
\fi%
}

\makeatother

\begin{document}

% format
%\baselineskip=18pt  % a la harvmac
%\numberwithin{equation}{section}  % make eq labels (sec.num)
%\allowdisplaybreaks  % allow page breaks in displayed eqs

% print date, time and filename 
%\pagestyle{myheadings}
%\markright{{\tt \jobname.tex} -- \today{} \now}

%%%%%%%%%%%%%%%%%%%%%%%%%%%%%%%%%%%%%%%%%%%
%%%        TITLE BEGINS HERE
%%%%%%%%%%%%%%%%%%%%%%%%%%%%%%%%%%%%%%%%%%%

%% ========== title (note version) begins here ==========
%
%\vspace*{-1cm}
%\begin{center}
% {\Large\bf Title of the Document}
%\end{center}
%\vspace*{-.5cm}
%
%% ========== title (note version) ends here ==========

%% ========== title (paper version, a la harvmac) begins here ==========

\preprint{EFI-10-26}

\title{Hypercharge Flux, Exotics, and Anomaly Cancellation in F-theory GUTs}

\author{Joseph Marsano} 
%\email{marsano@uchicago.edu}

\affiliation{Enrico Fermi Institute, University of Chicago\\ 5640 S Ellis Ave, Chicago, IL 60637, USA}

\begin{abstract}
We sharpen constraints related to hypercharge flux in F-theory GUTs that possess $U(1)$ symmetries and argue that they arise as a consequence of 4-dimensional anomaly cancellation.  This gives a physical explanation for all restrictions that were observed in spectral cover models while demonstrating that the phenomenological implications for a well-motivated set of models are not tied to any particular formalism.
\end{abstract}

\maketitle

%\newpage
%\setcounter{page}{1} % don't number title page

%% ========== title (paper version, a la harvmac) ends here ==========

%%%%%%%%%%%%%%%%%%%%%%%%%%%%%%%%%%%%%%%%%%%
%%%           TITLE ENDS HERE
%%%%%%%%%%%%%%%%%%%%%%%%%%%%%%%%%%%%%%%%%%%

%\tableofcontents
%\printindex

%%%%%%%%%%%%%%%%%%%%%%%%%%%%%%%%%%%%%%%%%%%
%%%        MAIN TEXT BEGINS HERE
%%%%%%%%%%%%%%%%%%%%%%%%%%%%%%%%%%%%%%%%%%%

\section{Introduction}

The vastness of the string landscape presents a serious obstacle for studying particle physics in string theory.  To make progress, it is often helpful to adopt a bottom-up approach \cite{Aldazabal:2000sa} that mirrors the successful techniques of effective field theory.  Type II  string theories provide a natural setting for this since the charged degrees of freedom can localize on branes that probe only a small part of the compactification geometry.  The low energy physics associated to these branes is captured by a non-Abelian gauge theory whose bare coupling constants at the compactification scale are determined by local geometric data.

This approach is particularly appealing for the construction of Grand Unified Theories (GUTs) \cite{Donagi:2008ca,Beasley:2008kw,Hayashi:2008ba} as the charged sector is engineered on a single stack of branes.   The volume of the internal cycle wrapped by the branes introduces a new scale into the problem that can help to realize the small observed hierarchy between $M_{\rm GUT}$ and $M_{\text{Planck}}$.
%The volume of the internal cycle that is wrapped by these branes introduces a new scale into the problem that can help to realize the small observed hierarchy between the GUT and Planck scales.
In this setting, the large top Yukawa coupling suggests an underlying exceptional group structure \cite{Tatar:2006dc} that motivates the study of nonperturbative type II configurations described by M-theory or F-theory.  The latter has received significant attention over the past few years in large part because powerful techniques of algebraic geometry are available to simplify the analysis.  

Most approaches to F-theory GUTs make crucial use of two important ingredients.  The first is the presence of $U(1)$ symmetries that can be used to protect against proton decay \cite{Tatar:2006dc,Marsano:2009gv,Blumenhagen:2009yv,Marsano:2009wr,Grimm:2009yu} or to motivate scenarios for how supersymmetry breaking is mediated to the Standard Model \cite{Marsano:2008jq}.
%, which typically originate from some underlying $E_8$ structure that is broken down to $SU(5)_{\rm GUT}$.  Symmetries of this type can be used to protect against proton decay \cite{Tatar:2006dc,Marsano:2009gv,Blumenhagen:2009yv,Marsano:2009wr,Grimm:2009yu} as well as to motivate scenarios for how supersymmetry breaking is mediated to the Standard Model \cite{Marsano:2008jq}.  
The second important ingredient is ``hypercharge flux", which provides an elegant mechanism for breaking the GUT group while addressing the doublet-triplet splitting problem \cite{Beasley:2008kw}.  In explicit constructions based on spectral cover techniques \cite{Donagi:2009ra}, these two ingredients appear to be interrelated \cite{Marsano:2009gv,Marsano:2009wr}; spectral cover models with a particular set of $U(1)$ symmetries tend to exhibit tight constraints on how ``hypercharge flux" can be distributed among the matter curves where charged fields localize \cite{Marsano:2009gv}.  This, in turn, has a striking impact on the 4-dimensional physics of all F-theory GUT models built to date.

The goal of this letter is to understand the nature and source of these constraints.  Because of the dramatic phenomenological implications \cite{Marsano:2009gv}, it is crucial to understand if the relationship between $U(1)$ symmetries and ``hypercharge flux" represents a limitation of our current model-building toolbox or a more general lesson with an intrinsic physical origin.  One indication of the latter can be found in a recent paper of Dudas and Palti \cite{Dudas:2010zb}, who noticed a simple pattern in the distribution of ``hypercharge flux" in a set of spectral cover models.  It is not hard to prove their relations for generic (suitably nondegenerate) spectral cover models and we do this in the upcoming paper \cite{MSSNinprogress}.  More intriguing, however, is that we can rewrite the original Dudas-Palti observation in a simple way that does not make explicit reference to spectral covers at all
%The goal of this letter is to understand the nature and source of these constraints.  
%Because of their dramatic phenomenological implications \cite{Marsano:2009gv}, it is crucial to determine if they carry an intrinsic physical meaning or represent artifacts of a particular model-building formalism as suggested in \cite{Cecotti:2010bp}.  One indication of the former can be found in a recent paper of Dudas and Palti \cite{Dudas:2010zb}, who noticed a simple pattern in the distribution of ``hypercharge flux" in a set of spectral cover models.  It is not hard to prove their relations for generic spectral cover models and we do this in the upcoming paper \cite{MSSNinprogress}.  More intriguing, however, is that the observation of Dudas and Palti can be written in a simple manner that does not make explicit reference to spectral covers at all
\begin{equation}\sum_{\substack{\mathbf{10}\text{ matter}\\ \text{curves, }a}} q_a\int_{\Sigma_{\mathbf{10}}^{(a)}}\omega_Y = \sum_{\substack{\mathbf{\overline{5}}\text{ matter}\\ \text{curves, }i}} q_i\int_{\Sigma_{\mathbf{\overline{5}}}^{(i)}}\omega_Y\label{DP}\end{equation}
Here, $q_a$ denotes the common $U(1)$ charge of $\mathbf{10}$ or $\mathbf{\overline{5}}$ fields that localize along curves $\Sigma^{(x)}$ in the compactification and $\omega_Y$ is a ``hypercharge flux" that is chosen to ensure that the $U(1)_Y$ gauge boson remains massless.  A relation this simple should have a physical origin and, in this letter, we will demonstrate that it is a consequence of 4-dimensional anomaly cancellation.  
In addition to clarifying the physics of all known constraints of spectral cover models, this observation allows us to derive a generalization of \eqref{DP} that must be satisfied by \emph{any} F-theory GUT that combines $U(1)$ symmetries and ``hypercharge flux" regardless of how it is constructed.
%, including models with a type IIB orientifold limit, that combines $U(1)$ symmetries and ``hypercharge flux".
% to break $SU(5)_{\rm GUT}$.
%clarifies the physics of all restrictions that have appeared in spectral cover constructions and demonstrates how they generalize to F-theory GUTs.
Among the many implications for phenomenology, our results imply that any $U(1)$ symmetry in a model that combines ``hypercharge flux" with the flavor scenario of \cite{Heckman:2008qa} must be $U(1)_{B-L}$, which cannot address $\mu$ or dimension 5 proton decay.  Insisting on the existence of a $U(1)_{PQ}$ symmetry to deal with these necessarily introduces charged exotics into the spectrum.
% can only be combined  the existence of charged exotics in any of a set of phenomenologically-motivated F-theory GUTs that combine the flavor scenario of \cite{Heckman:2008qa} with ``hypercharge flux" and the existence of a $U(1)_{PQ}$ symmetry.

\section{F-theory GUTs and Anomaly Cancellation}

\subsection{Spectrum and ``Hypercharge Flux"}

The charged sector of an F-theory GUT model is described by the 8-dimensional worldvolume theory that describes the physics of a stack of 7-branes.  This theory, which we take to have gauge group $SU(5)_{\rm GUT}$, is compactified on a complex surface $S_{\rm GUT}$ and can be UV completed by embedding that surface into a consistent F-theory compactification.  Adjoint-valued fields propagate throughout the 8-dimensional worldvolume but the model contains additional degrees of freedom in the $\mathbf{10}$ and $\mathbf{\overline{5}}$ representations (and their conjugates) that localize on holomorphic "matter curves" in $S_{\rm GUT}$.  Determining the 4-dimensional spectrum requires a dimensional reduction in either case and can be influenced by introducing suitable fluxes into the model.

While most of these fluxes descend from the bulk of the compactification, worldvolume flux plays an important role.  An internal flux of the $U(1)_Y$ gauge field can break $SU(5)_{\rm GUT}$ down to the MSSM gauge group and, when chosen correctly, remove unwanted degrees of freedom like Higgs triplets and leptoquarks \cite{Beasley:2008kw}.  In general, the net chirality of leptoquarks that descend from the $SU(5)_{\rm GUT}$ adjoint is determined by an index theorem \cite{Beasley:2008kw}
\begin{equation*}n_{(\mathbf{3},\mathbf{2})_{-5/6}}-n_{(\mathbf{\overline{3}},\mathbf{2})_{+5/6}}=\int_{S_{\rm GUT}} c_1(S_{\rm GUT})\wedge c_1(L_Y^{5/6})\end{equation*}
where $L_Y$ is a line bundle that specifies the ``hypercharge flux".  The spectrum on a matter curve $\Sigma$, on the other hand, is computed as \cite{Beasley:2008kw}
\begin{equation*}n_R-n_{\overline{R}}=\int_{\Sigma}\! c_1\left(V_{\Sigma}\otimes L_Y^{Y_R}\right)
=\int_{\Sigma} \left[c_1(V_{\Sigma})+M_{\Sigma}c_1(L_Y^{Y_R})\right]
\end{equation*}
where $V_{\Sigma}$ is a bundle of rank $M_{\Sigma}$ that roughly encodes the "bulk" fluxes and $Y_R$ is the $U(1)_Y$ charge of fields in the representation $R$.  The bundle $V_{\Sigma}$ and its rank $M_{\Sigma}$ are intrinsic properties of the matter curve $\Sigma$ but the charges $Y_R$ can differ among the various MSSM multiplets contained in the $SU(5)_{\rm GUT}$ multiplet that localizes there.  In this way, a nontrivial ``hypercharge flux" can be used to generate incomplete GUT-multiplets, which is very useful for obtaining Higgs doublets without their triplet partners.  The ranks $M_{\Sigma}$ are all 1 for spectral cover models that are suitably nondegenerate but can be larger in more general constructions \cite{Hayashi:2008ba,Cecotti:2010bp}.

\subsection{Constraints on ``Hypercharge Flux" from MSSM Gauge Anomalies}

When building models, we need some freedom to distribute ``hypercharge flux" among the matter curves that are present.  This freedom must be limited, though, because ``hypercharge flux" induces a chiral spectrum with respect to the MSSM gauge groups that generically leads to anomalies.  The $SU(3)^3$ anomaly, for instance, is proportional to
\begin{equation*}\begin{split}3\!\!\!\! \sum_{\substack{\mathbf{10}\text{ matter}\\ \text{curves, }i}}\!\!\!\!\!\! M_{\Sigma_{\mathbf{10}}^{(i)}}&\int_{\Sigma_{\mathbf{10}}^{(i)}}\!\!\!\! c_1(L_Y)\,\,\, - \!\!\sum_{\substack{\mathbf{\overline{5}}\text{ matter }\\\text{curves, }a}}\!\!\!\!\!\! M_{\Sigma_{\mathbf{\overline{5}}}^{(a)}}\int_{\Sigma_{\mathbf{\overline{5}}}^{(a)}}\!\!\!\!c_1(L_Y) \\
&+ 5\int_{S_{\rm GUT}}\!\!\!\!\!\! c_1(S_{\rm GUT})\wedge c_1(L_Y)\end{split}\end{equation*}
Since this must cancel regardless of how we choose $c_1(L_Y)$, we see that the matter curves of any consistent F-theory GUT model should satisfy
\begin{equation*}3\sum_{\substack{\mathbf{10}\text{ matter}\\ \text{curves, }i}}M_{\Sigma_{\mathbf{10}}^{(i)}}[\Sigma_{\mathbf{10}}^{(i)}]-\sum_{\substack{\mathbf{\overline{5}}\text{ matter}\\ \text{curves, }a}}M_{\Sigma_{\mathbf{\overline{5}}}^{(a)}}[\Sigma_{\mathbf{\overline{5}}}^{(a)}] + 5[c_1]=0\label{SU33}\end{equation*}
where $[c_1]$ is the anti-canonical curve of $S_{\rm GUT}$.  This relation is well-known \cite{Andreas:2009uf,Donagi:2009ra} for constructions with $M_{\Sigma_{\mathbf{10}}^{(i)}}=M_{\Sigma_{\mathbf{\overline{5}}}^{(a)}}=1$ and has been derived using a ``stringy" anomaly cancellation argument \cite{Donagi:2009ra}.  It is amusing to see, however, that it can be understood already as a consequence of anomaly cancellation in 4-dimensions.

Cancellation of mixed gauge anomalies involving $U(1)_Y$ is not guaranteed for generic choices of $L_Y$ because, in most cases, the hypercharge gauge boson is lifted through an induced coupling to RR fields \cite{Beasley:2008kw}.  The conditions that $L_Y$ must satisfy in order to prevent this are known in F-theory and correspond to constructing $L_Y$ from a  $(1,1)$-form, $\omega_Y\sim c_1(L_Y)$, that trivializes in the full compactification.  Any ``hypercharge flux" of this type will necessarily be constrained; at the very least, its distribution among the matter curves must guarantee that all MSSM gauge anomalies are cancelled.  This leads to the conditions
\begin{equation}\begin{split}0 = \sum_{\substack{\mathbf{10}\text{ matter}\\ \text{curves, }i}}\!\!\!\!\!\! M_{\Sigma_{\mathbf{10}}^{(i)}}\int_{\Sigma_{\mathbf{10}}^{(i)}}\!\! c_1(L_Y) = \sum_{\substack{\mathbf{\overline{5}}\text{ matter}\\ \text{curves, }a}}\!\!\!\!\!\! M_{\Sigma_{\mathbf{\overline{5}}}^{(a)}}\int_{\Sigma_{\mathbf{\overline{5}}}^{(a)}}\!\! c_1(L_Y)
\end{split}\label{MSSManomalies}\end{equation}
that are easy to verify in generic F-theory GUT models \cite{Andreas:2009uf} with a massless $U(1)_Y$.

\subsection{Implications of Mixed Gauge Anomalies}

We would now like to ask if a ``hypercharge flux" $\omega_Y$ that doesn't lift $U(1)_Y$ exhibits any additional properties in a geometry that engineers bulk $U(1)$ symmetries in addition to $SU(5)_{\rm GUT}$ \cite{U1footnote}.  To address this, let us consider what happens when we turn on this flux \emph{and no other fluxes}.  
%To investigate the properties of $\omega_Y$ in these models, let us use it to construct a flux that is purely in the $U(1)_Y$ direction and consider what happens when we turn on this flux \emph{and nothing else}.  
Our flux will induce a nontrivial spectrum but, because all $U(1)$'s remain massless, it cannot give rise to any gauge anomalies \cite{Fluxfootnote}.
%{\footnote{A subtlety arises here because some F-theory compactifications cannot be globally consistent unless a bulk $G$-flux is added \cite{Witten:1996md}.  
%The fields $\hat{C}_0$ and $\hat{C}_2$ are related by self-duality of $C_4$ and their 4-dimensional couplings descend from the bulk interaction $\int\,C_4\wedge G\wedge G$.  
%Because $\omega_Y$ is globally trivial, it cannot contribute to these couplings and hence cannot contribute to the anomalies, either.}}.
%does not play any role in the emergence of these couplings from dimensional reduction so they must be independent of the rescaling $N$, along with any anomalies that they cancel.}}.
Of particular interest to us are mixed anomalies with insertions of both MSSM and $U(1)$ currents since these only get contributions from the chiral fields that localize on matter curves in $S_{\rm GUT}$.  We will see that the Dudas-Palti relations \eqref{DP} for spectral cover models simply express a set of nontrivial relations that the $(1,1)$-form $\omega_Y$ must satisfy in order for these 4-dimensional mixed gauge anomalies to cancel.

To make things completely explicit, we use $\omega_Y$ to define a line bundle ${\cal{L}}_Y$ on the GUT 7-branes that defines a nontrivial $U(1)_Y$ background.  We further normalize that background so that all charged fields on matter curves are sections of the integer quantized gauge bundles listed below 
\begin{equation}\begin{array}{c|c|c}SU(5)
& SU(3)\times SU(2)\times U(1)_Y
& \text{Bundle} \\ \hline
\mathbf{10} & (\mathbf{1},\mathbf{1})_{+1} & {\cal{L}}_Y^6\\
& (\mathbf{3},\mathbf{2})_{+1/6} & {\cal{L}}_Y \\
& (\mathbf{\overline{3}},\mathbf{1})_{-2/3} & {\cal{L}}_Y^{-4} \\ \hline
\mathbf{\overline{5}} & (\mathbf{\overline{3}},\mathbf{1})_{+1/3} & {\cal{L}}_Y^2 \\
& (\mathbf{1},\mathbf{2})_{-1/2} & {\cal{L}}_Y^{-3}
\end{array}\end{equation}

We now determine the contributions to mixed gauge anomalies that arise from the chiral spectrum on a generic $\mathbf{10}$ or $\mathbf{\overline{5}}$ matter curve.  To obtain \eqref{DP} and its generalization beyond spectral cover models, it will be sufficient to consider anomalies of the type $G_{SM}^2\times U(1)$, where $G_{SM}$ denotes a Standard Model gauge group.

Consider first the contribution from fields that localize on a $\mathbf{10}$ curve, $\Sigma_{\mathbf{10}}^{(a)}$, which carry a $U(1)$ charge $q_a$.  Denoting the $M_{\Sigma_{\mathbf{10}}^{(a)}}$-weighted $U(1)_Y$ flux there by $N_a$
\begin{equation}N_a = M_{\Sigma_{\mathbf{10}}^{(a)}}\int_{\Sigma_{\mathbf{10}}^{(a)}}c_1({\cal{L}}) = M_{\Sigma_{\mathbf{10}}^{(a)}}\int_{\Sigma_{\mathbf{10}}^{(a)}}\omega_Y\end{equation}
the contributions to mixed $G_{SM}^2\times U(1)$ anomalies are
\begin{equation*}\begin{array}{c|c|c|c|c}
\text{Multiplet} & \text{Chir} & SU(3)^2 U(1) & SU(2)^2 U(1) & U(1)_Y^2 U(1) \\ \hline
(\mathbf{1},\mathbf{1})_{+1} & 6N_a & 0 & 0 & 6q_aN_a \\
(\mathbf{3},\mathbf{2})_{+1/6} & N_a & 2q_aN_a & 3q_aN_a & q_aN_a/6 \\ 
(\mathbf{\overline{3}},\mathbf{1})_{-2/3} & -4N_a & -4q_aN_a & 0 & -16q_aN_a/3 \\ \hline
\text{Total} & &-2q_aN_a & 3q_aN_a & 5q_aN_a/6 \\
\end{array}\end{equation*}
Note that a negative chirality means that we obtain zero modes of the conjugate multiplet, which carry an opposite $U(1)$ charge.  We now do the same thing for fields on a $\mathbf{\overline{5}}^{(i)}$ curve that carry $U(1)$ charge $q_i$.  Letting $N_i$ denote the $M_{\Sigma_{\mathbf{\overline{5}}^{(i)}}}$-weighted $U(1)_Y$ flux
\begin{equation}N_i = M_{\Sigma_{\mathbf{\overline{5}}^{(i)}}}\int_{\Sigma_{\mathbf{\overline{5}}^{(i)}}}c_1({\cal{L}}) = M_{\Sigma_{\mathbf{\overline{5}}^{(i)}}}\int_{\Sigma_{\mathbf{\overline{5}}^{(i)}}}\omega_Y\end{equation}
we find
\begin{equation*}\begin{array}{c|c|c|c|c} 
\text{Multiplet} & \text{Chir} & SU(3)^2 U(1) & SU(2)^2 U(1) & U(1)_Y^2 U(1) \\ \hline
(\mathbf{\overline{3}},\mathbf{1})_{+1/3} & 2N_i & 2q_iN_i & 0 & 2q_iN_i/3 \\
(\mathbf{1},\mathbf{2})_{-1/2} & -3N_i & 0 & -3q_iN_i & -3q_iN_i/2 \\ \hline
\text{Total} & & 2q_iN_i & -3q_iN_i & -5q_iN_i/6
\end{array}\end{equation*}

From this, we see that cancellation of all $G_{SM}^2\times U(1)$ anomalies implies that $\omega_Y$ must satisfy
\begin{equation}\sum_{\substack{\mathbf{10}\text{ matter}\\ \text{curves,}a}}\!\!\!\! q_a M_{\Sigma_{\mathbf{10}}^{(a)}}\int_{\Sigma_{\mathbf{10}}^{(a)}}\omega_Y = \sum_{\substack{\mathbf{\overline{5}}\text{ matter}\\ \text{curves,}i}}\!\!\!\! q_i M_{\Sigma_{\mathbf{\overline{5}}}^{(i)}}\int_{\Sigma_{\mathbf{\overline{5}}}^{(i)}}\omega_Y\label{gDP}\end{equation}
which, for $M_{\Sigma_{\mathbf{10}}^{(a)}}=M_{\Sigma_{\mathbf{\overline{5}}}^{(i)}}=1$, is nothing other than the Dudas-Palti relations \eqref{DP}.  We refer to \eqref{gDP} as the generalized Dudas-Palti relations, which must hold for any $\omega_Y$ that can be used to construct ``hypercharge flux" in an $SU(5)_{\rm GUT}$ F-theory GUT model with an extra $U(1)$ symmetry.  It is easy to see that other mixed anomalies, as well as the $U(1)^3$ anomaly, vanish without giving rise to any additional constraints.  Though the story is less constrained than in 6-dimensions \cite{Kumar:2009ac}, it would be interesting to pursue a more general analysis of anomaly cancellation in 4-dimensional F-theory compactifications in the future.

\section{Implications of the Generalized Dudas-Palti Relations}

The first question to ask about \eqref{gDP} and \eqref{MSSManomalies} is whether they represent all of the nontrivial constraints on the distribution of ``hypercharge flux" in F-theory GUTs.  In the case of spectral cover models, we suspect that they do because it appears that one can use spectral covers to construct, at least in principle, all distributions of ``hypercharge flux" that satisfy them \cite{MSSNinprogress}.  Based on this, it is natural to conjecture that, even for more general classes of F-theory GUTs,  \eqref{gDP} and \eqref{MSSManomalies} represent the only constraints.
%even for more general classes of F-theory GUTs but we do not have a proof of this statement.

In light of this, we should correct some misstatements that were made in \cite{Marsano:2009gv}.  There, it was claimed that the presence of ``hypercharge flux" on $\mathbf{\overline{5}}$ matter curves automatically implied that ``hypercharge flux" must thread some $\mathbf{10}$ matter curves as well.  The DP relations \eqref{DP} do not forbid a configuration in which ``hypercharge flux" threads only $\mathbf{\overline{5}}$ curves, though, and it is possible to construct spectral covers that do precisely this \cite{MSSNinprogress}.

Finally, let us comment on implications of the generalized Dudas-Palti relations \eqref{gDP} for F-theory model building.  While several approaches to flavor have been suggested in the past few years \cite{Font:2008id}, the mechanism of wave function overlaps is particularly attractive \cite{Heckman:2008qa}.  This mechanism requires all three generations of the $\mathbf{10}$ to localize on one matter curve and similar for all three generations of the $\mathbf{\overline{5}}$.  The Higgs fields then lie on distinct matter curves, $\Sigma_{\mathbf{\overline{5}}}^{(H_u)}$ and $\Sigma_{\mathbf{\overline{5}}}^{(H_d)}$, which must have $M_{\Sigma_{\mathbf{\overline{5}}}^{(H_u)}}=M_{\Sigma_{\mathbf{\overline{5}}}^{(H_d)}}=1$ and carry +1 and -1 units of ``hypercharge flux", respecitvely, to lift the triplets \cite{Beasley:2008kw}.  Crucial to this scenario is that ``hypercharge flux" not be allowed to thread any curve $\Sigma$ other than $\Sigma_{\mathbf{\overline{5}}}^{(H_u)}$ and $\Sigma_{\mathbf{\overline{5}}}^{(H_d)}$; if it did, we would obtain massless matter fields on $\Sigma$ that do not comprise a complete GUT multiplet.  As one assumes that the Standard Model fields are engineered as complete GUT multiplets, the threading of ``hypercharge flux" through such a $\Sigma$ necessarily introduces new charged exotics into the spectrum \cite{Marsano:2009gv}.

%To realize this mechanism,
%one would like to engineer all three generations of the $\mathbf{10}$ on one matter curve and all three generations of the $\mathbf{\overline{5}}$ on a second matter curve.  The Higgs fields then lie on distinct matter curves, $\Sigma_{\mathbf{\overline{5}}}^{(H_u)}$ and $\Sigma_{\mathbf{\overline{5}}}^{(H_d)}$, which must have $M_{\Sigma_{\mathbf{\overline{5}}}^{(H_u)}}=M_{\Sigma_{\mathbf{\overline{5}}}^{(H_d)}}=1$ and carry +1 and -1 units of ``hypercharge flux", respectively, to lift the triplets \cite{Beasley:2008kw}.  Crucial to this scenario is that ``hypercharge flux" not be allowed to thread any curve $\Sigma$ other than $\Sigma_{\mathbf{\overline{5}}}^{(H_u)}$ and $\Sigma_{\mathbf{\overline{5}}}^{(H_d)}$; if it did, we would obtain massless matter fields on $\Sigma$ that do not comprise a complete GUT multiplet.  As one assumes that the standard model fields are engineered as complete GUT multiplets, the threading of ``hypercharge flux" through such a $\Sigma$ 
%will necessarily introduce new charged exotics into the spectrum \cite{Marsano:2009gv}.

If we wish to combine this scenario with a $U(1)$ symmetry, the generalized Dudas-Palti relations \eqref{gDP} imply that the charges $q_{H_u}$ and $q_{H_d}$ associated to the matter curves $\Sigma_{\mathbf{\overline{5}}^{(H_u)}}$ and $\Sigma_{\mathbf{\overline{5}}^{(H_d)}}$ must satisfy
\begin{equation}q_{H_u}-q_{H_d}=0\label{Hcurvecharges}\end{equation}
The doublet $H_u$ comes from a $\mathbf{5}$ rather than a $\mathbf{\overline{5}}$, though, so its charge is actually $-q_{H_u}$.  Writing \eqref{Hcurvecharges} in terms of the actual $H_u$ and $H_d$ charges we get
\begin{equation}Q(H_u) + Q(H_d)=0\label{Hcharges}\end{equation}
What type of $U(1)$ symmetry can this be?  Because all $\mathbf{10}$'s ($\mathbf{\overline{5}}$'s) are engineered on a single curve, all of them must carry a common charge.  The only $U(1)$ symmetry of this type that commutes with $SU(5)$, satisfies \eqref{Hcharges}, and preserves the MSSM superpotential is the famous $U(1)_{\chi}$, which is the linear combination of $U(1)_Y$ and $U(1)_{B-L}$ that enters naturally in $SO(10)$ unification models.  We see that $PQ$ symmetries, broadly defined as $U(1)$'s for which \eqref{Hcharges} does not hold, cannot be combined with the desired distribution of hypercharge flux.  If we insist on realizing all 3 generations of $\mathbf{10}$'s ($\mathbf{\overline{5}}$'s) on a single matter curve, the presence of $U(1)_{PQ}$ implies the existence of additional charged matter fields that do not come in complete GUT multiplets \cite{Marsano:2009gv}.

\paragraph*{Acknowledgements:} I am grateful to N.~Saulina and S.~Sch\"afer-Nameki for valuable discussions during the course of this work and many enjoyable collaborations on the study of F-theory GUTs.  I am also grateful to S.~Sethi for encouragement and helpful discussion as well as S.~Cecotti, C.~Cordova, J.~Heckman, and C.~Vafa for explaining a crucial aspect of their work \cite{Cecotti:2010bp}.  I thank the Physics Department at The Ohio State University and the organizers of the String Vacuum Project 2010 Fall meeting for their hospitality.
% during the final stages of this work.  
This research is supported by DOE grant DE-FG02-90ER-40560 and NSF grant PHY-0855039.

\bibliographystyle{revtex}
\renewcommand{\refname}{Bibliography}
\addcontentsline{toc}{section}{Bibliography}

\providecommand{\href}[2]{#2}\begingroup\raggedright\endgroup

\end{document}